# Infections Forecasting and Intervention Effect Evaluation for COVID-19 via a Data-Driven Markov Process and Heterogeneous Simulation

**Quan-Lin Li[1,\*]; Chengliang Wang[1]; Yiming Xu[1]; Chi Zhang[1,\*]; Yanxia Chang[1]; Xiaole Wu[2]; Zhen-Ping Fan[3]; Zhi-Guo Liu[4]**

**Abstract:** The Coronavirus Disease 2019 (COVID-19) pandemic has caused tremendous amount of deaths and a devastating impact on the economic development all over the world. Thus, it is paramount to control its further transmission, for which purpose it is necessary to find the mechanism of its transmission process and evaluate the effect of different control strategies. To deal with these issues, we describe the transmission of COVID-19 as an explosive Markov process with four parameters. The state transitions of the proposed Markov process can clearly disclose the terrible explosion and complex heterogeneity of COVID-19. Based on this, we further propose a simulation approach with heterogeneous infections. Experimentations show that our approach can closely track the real transmission process of COVID-19, disclose its transmission mechanism, and forecast the transmission under different non-drug intervention strategies. More importantly, our approach can helpfully develop effective strategies for controlling COVID-19 and appropriately compare their control effect in different countries/cities.

[1]School of Economics and Management, Beijing University of Technology, Beijing 100124, China. [2]School of Management, Fudan University, Shanghai 200433, China. [3]The Fifth Department of Cadre Health Care,The Second Medical Center & National Clinical Research Center for Geriatric Diseases, Chinese PLA General Hospital. [4]The Department of Blood Transfusion,The Fifth Medical Center, Chinese PLA General Hospital.

\*✉e-mail: liquanlin@tsinghua.edu.cn   ✉e-mail: chizhang@live.com

On January 24, 2020, a Chinese Lunar New Year eve, the Coronavirus Disease 2019 (COVID-19) outbreak began in Wuhan, a beautiful provincial capital city in Hubei province, China. Up to now, the COVID-19 has become even more serious all over the world. As of Dec. 29, 2020, there are more than 81 million cumulative cases in the world, including more than 22 million active cases (https://github.com/CSSEGISandData/COVID-19). The global outbreak transmission of COVID-19 has seriously endangered human health, extremely destroyed daily life of people, and caused huge economic losses in many countries[1,2]. Therefore, it is imperative to find some effective methods for evaluating and predicting the transmission processes of COVID-19 and the control effect of intervention strategies. For the evaluation and prediction of COVID-19, the current mainstream is the SIR based epidemiological model[3-5]. The classical SIR model ignores several important factors, such as the heterogeneity and explosibility of COVID-19, and a huge initial number of patients. Furthermore, a patient can infect multiple persons each time. Therefore, such SIR models are not suitable for the explosive growth of patients, and the inter-person infection among a huge initial number of patients.

Now, it is necessary to explain some key characteristics of COVID-19 transmission processes. The transmission of COVID-19 consists of multiple stages with different levels of infection[6,7]. The exponential growth of infections was found during its initial stage[8-10]. COVID-19 is observed to be highly infectious[11,12] with a high portion of severe cases[2]. In addition, some studies used the Poisson processes to model the COVID-19 transmission[13], and provided the Markov Chain Monte Carlo method to evaluate the basic reproductive number[14]. The transmission of COVID-19 was found to be seasonal[15,16]. Also, the infection of COVID-19 becomes really terrible at a very low temperature in the winter[17]. In the beginning of 2021, the COVID-19 has become even more active and serious in many countries. Therefore, the prevention and control of COVID-19 in the world will continue facing huge challenges in 2021.

So far, we have found that the prevention and control of COVID-19 include four non-drug intervention strategies[18-23]: Travel restrictions[24-27], early identification and isolation[28-36], increasing social distance[37-40], and exposure restrictions[30,41-44]. It was indicated that we will have to live with COVID-19 in the decades ahead[45]. Thus, the non-drug intervention strategies can play an important role in the prevention and control of COVID-19 transmission due to the fact that no effective drug is found for treating COVID-19.

To describe and analyze the complex transmission of COVID-19, this paper is devoted to developing an approach for evaluating and predicting the transmission processes of COVID-19, which is modeled as a Markov process with explosive structure. Our model needs four basic parameters to capture the basic characteristics of the COVID-19 transmission process: the explosive growth of patients, and the batching infection of an infected person. These four parameters can be estimated based on public data (https://github.com /CSSEGISandData/COVID-19). We also propose a simulation approach with heterogeneous structure based on the proposed Markov model and the four parameters. Experimentations on the cases of six countries/cities show that our approach works well in tracking the real transmission process of COVID-19.

## Methods

We set up a novel block-structured Markov process to describe and analyze the virus generation, infection and large-scale diffusion of COVID-19. At the same time, this Markov process can explicitly express and track the comprehensive control effect of the non-drug intervention strategies in different time intervals on COVID-19 transmission.

Our Markov process of COVID-19 captures several key random factors in the transmission process of COVID-19. By applying the theory of Markov processes, the random factors can be determined based on the framework of exponential distributions and/or Poisson processes. Then, the random factors can be further simplified to four key parameters as follows:

1. The infection rate $\lambda$: The number of batches of healthy people who can be infected by one infected person per unit time. Note that the time interval between two successively infected batches of patients is exponential with infection rate $\lambda$.

2. The disappearing rate $\mu$: The number of disappearing patients because of being cured or death per unit time. Note that the time interval between two successively disappearing patients is exponential with disappearing rate $\mu$.

3. The infection batch size $d$: The number of healthy people that a patient can infect each time.

4. The initial patient number $k$: The number of patients at the initial time of our observation in a chosen area (for example, a country, a city or a village).

Let $N(t)$ be the number of COVID-19 patients at time $t \geq 0$. We depict the state transition relation of the Markov system $\{N(t): t \geq 0\}$ in Fig. 1.

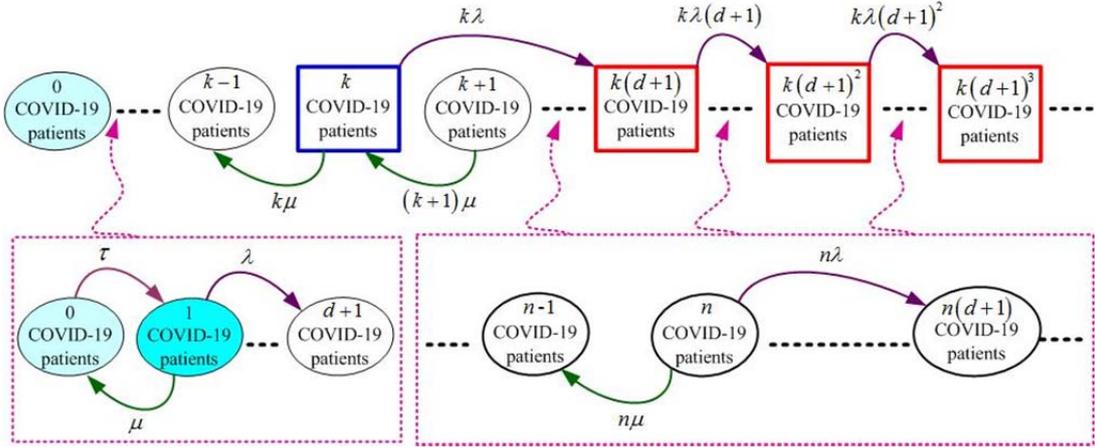

**Figure 1. The state transition relations of the Markov process of COVID-19. $\tau$ denotes the initial infection rate that COVID-19 is found for the first time. In general, $\tau$ is very small.**

From Fig. 1, we define the sets of states as follows:

Level 0: $L_0 = \{0, 1, 2, \dots, k-1\}$;

Level 1: $L_1 = \{k, k+1, k+2, \dots, k(d+1) - 1\}$;

Level 2: $L_2 = \{k(d+1), k(d+1) + 1, k(d+1) + 2, \dots, k(d+1)^2 - 1\}$;

Level $n$: $L_n = \{k(d+1)^{n-1}, k(d+1)^{n-1} + 1, k(d+1)^{n-1} + 2, \dots, k(d+1)^n - 1\}, n \geq 3$.

Using the different levels, the Markov process $\{N(t): t \geq 0\}$ is a QBD (quasi-birth and death) process whose infinitesimal generator is given by

$$Q = \begin{pmatrix} A_{0,0} & A_{0,1} & & & \\ A_{1,0} & A_{1,1} & A_{1,2} & & \\ & A_{2,1} & A_{2,2} & A_{2,3} & \\ & & \ddots & \ddots & \ddots \end{pmatrix},$$

where $A_{i,j}$ is a matrix of suitable size based on the sizes of the above levels from Level $i$ to Level $j$.

Let
$$\boldsymbol{\omega} = (0, 0, \ldots, 0, 1, 0, 0, \ldots),$$
where 1 is at the $(k+1)$st position of vector $\boldsymbol{\omega}$, while all the other elements are zero. We define
$$p_n(t) = P\{N(t) = n\}, \quad n = 0, 1, 2, \ldots,$$
$$\mathbf{P}(t) = (p_0(t), p_1(t), p_2(t), p_3(t), \ldots).$$

From the Chapman-Kolmogorow equations, it is easy to see that
$$\mathbf{P}(t) = \boldsymbol{\omega} \exp\{Qt\}, \quad t \geq 0,$$
where $\exp\{Qt\}$ denotes $e^{Qt}$. Thus, the average number of COVID-19 patients at time $t$ is given by
$$E[N(t)] = \mathbf{P}(t) \times (0,1,2,3,\ldots)^T = \boldsymbol{\omega} \exp\{Qt\} \times (0,1,2,3,\ldots)^T, \quad t \geq 0,$$
where $(0,1,2,3,\ldots)^T$ is the transpose of the row vector $(0,1,2,3,\ldots)$.

On the one hand, the average number $E[N(t)]$ is a key index in the study of COVID-19 and closely related to the four basic parameters: $\lambda$, $\mu$, $d$, and $k$. On the other hand, it is worthwhile to note that the four parameters can be greatly influenced by the non-drug intervention strategies. Therefore, we need to analyze the relations between the average number $E[N(t)]$ and the non-drug intervention strategies by means of our Markov process of COVID-19.

Now, by using the daily reported data of COVID-19 (https://github.com/CSSEGISandData/COVID-19), we provide some effective methods to determine the four basic parameters: $\lambda$, $\mu$, $d$, and $k$. Further, our observation covers $m$ days. Let $n_i$ and $N_i$ be the numbers (or data) of newly infected patients of COVID-19 and all patients of COVID-19 on day $i$, respectively. We denote by $c_i$ the number of disappeared patients of COVID-19 on day $i$. Thus, the infection rate $\lambda$ from one patient of COVID-19 is given by

$$\lambda d = \frac{\frac{n_1}{N_1} + \frac{n_2}{N_2} + \frac{n_3}{N_3} + \cdots + \frac{n_{m-1}}{N_{m-1}} + \frac{n_m}{N_m}}{m}, \qquad (1)$$

where $d = 1, 2$, according to the actual data within a country or a city. Once $d$ is given, the infection rate $\lambda$ can be determined directly. Similarly, the disappearing rate $\mu$ of one patient of COVID-19 is given by

$$\mu = \frac{\frac{c_1}{N_1} + \frac{c_2}{N_2} + \frac{c_3}{N_3} + \cdots + \frac{c_{m-1}}{N_{m-1}} + \frac{c_m}{N_m}}{m}. \qquad (2)$$

Note that the initially patient number $k$ can be directly given from the actual data on the first day of our COVID-19 observation.

In order to take the value of the infection batch size, $d$, we can consider the patients as two groups with different batch sizes. Then, the overall batch size is the weighted average of these two groups. The weight values can be approximately designed according to the number of existing patients in the different groups. To better show this idea, let $k$ represents the number of the overall patients, and $k_i$ represents that of group $i$ ($i=1,2$). Then, the weight of group $i$ can be determined by $r_i = k_i / k$. See Fig. 1 for more details.

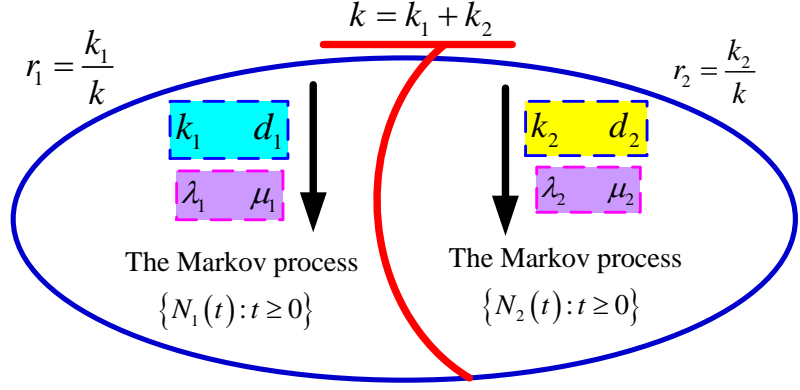

Figure 2. The different groups of COVID-19 transmission

Note that

$$E[N(t)] = E[N_1(t)] + E[N_2(t)]$$

We approximately infer the two weights, $r_i$, of the batch size of each group, and by adjusting their values, we can also make $E[N(t)]$ close to the real values.

## Results

In this section, we verify the effectiveness of our Markov method by simulating for several countries. Assuming that the parameters remain stable during the short-term observation period, we use the historical data in the observation period to predict the trend of active cases in the future. After that, the impact of three key parameters (i.e., the initially observed number $k$, the expansion batch size $d$ and the infection rate $\lambda$) on the transmission of COVID-19 is discussed. Note that the three parameters can be obviously influenced by the non-drug intervention strategies.

Our observation and analysis for the three parameters in the transmission process of COVID-19 is studied by means of our Markov processes of COVID-19; while some necessary and useful data of COVID-19 is taken up to the end of December, 2020, and the observed countries with the most active cases include the State of New York, India, Egypt, South Korea, Italy and Mexico and so forth (https://github.com/CSSEGISandData/COVID-19). Here, we can simulate and numerically analyze the evolutionary process of COVID-19 transmission for these countries, and discuss how the three parameters influence the average number $E[N(t)]$ of COVID-19 patients at time $t$. This allows us to evaluate the control effect of the non-drug intervention strategies on the COVID-19 transmission.

Now, we simulate the development of COVID-19 active cases in six countries/cities in a short-term period of 20 days. Different parameters are obtained according to the actual situation of State of New York, India, Egypt, South Korea, Italy and Mexico. Note that New York, India and Egypt represent a group of countries or regions without obvious intervention strategies, which can be seen from the relatively stable parameters during the observation period. We can also see that the parameters of the other countries have obvious change points, which indicates that they have taken obvious intervention strategies.

Because of the huge number of confirmed active cases in the United States, it may be better to study one of its representative states: New York State. According to the confirmed, deaths and cured cases in New York State from Nov. 6 to 25, 2020, as shown in Table 1, we obtained the numbers of active cases, the infection rates and the disappearing rates during this period. The average infection rate ($\lambda$) within these 20 days is determined according to Equation (1) as 0.035073852 and the average disappearing rate ($\mu$) within the same period is determined according to Equation (2) as 0.006084398. The initial number of active cases ($k$) at the beginning of our observation period (Nov. 6, 2020) is determined as 101592. We then simulated the development of active cases during this short-term period for New York State. In order for the simulated results to be close to the real data, we infer that the weights ($r$) of the two groups $d_1 = 1$, $d_2 = 2$ equal 0.971 and 0.029, respectively. The results are shown in Fig. 3.

We further simulated the growth processes of India in Asia (Nov. 1 to 20, 2020, Fig. 4) and Egypt in Africa (Nov. 1 to 20, 2020, Fig. 5). According to the data of India from Nov. 1 to 20, 2020, as shown in Table 2 and in Egypt from Nov. 1 to 20, 2020, as shown in Table 3, these simulations show that our model can well capture the transmission process of COVID-19 during the short-term observation period.

Different from the above three countries/states, the real data of South Korea, Italy and Mexico shows that there is a change point ($t_c$) in each number of their active cases, as shown in Fig. 6, 7 and 8. Once the change point is determined, we change the values of parameters, $\lambda$, $\mu$, and $d$, at that point. The existence of the change point reflects that the implementation of the non-drug intervention strategies has changed the COVID-19 transmission process. Therefore, we add a change point ($t_c$) during the simulation of South Korea, Italy, and Mexico.

First, based on Table 4, we calculated the four key parameters for the growth process of the COVID-19 active cases in South Korea from Nov. 2 to 21, 2020 and found that there is a change point ($t_c$ = 11) at which the values of these four parameters have changed. The reason behind this change is that, although South

Korea released the social distance policy, they relaxed the policy on Nov. 1, 2020. As a result, there is an uncontrollable rebound of the epidemic in winter. The parameters before $t_c$ are as follows: The average infection rate ($\lambda$) is 0.062135947 and the average disappearing rate ($\mu$) is 0.053096363. The initial number of active cases ($k$) at the beginning of our observation period (Nov. 2, 2020) is determined as 1825. We then simulated the development of active cases in South Korea during this period. In order to obtain simulated results close to the real data, we infer that the weights ($r$) of the two groups $d_1 = 1$, $d_2 = 2$ equal 0.940 and 0.060, respectively. The parameters after $t_c$ are as follows: The average infection rate is 0.09774732, the average disappearing rate is 0.038994525, and the weights ($r$) of the two groups $d_1 = 1$, $d_2 = 2$ equal 0.434 and 0.566, respectively. Fig. 6 shows the simulation results, versus with the real patient numbers of COVID-19.

For such countries with a change point, in addition to Korea, we also simulated Italy in Europe (Nov. 11 to 30, 2020, Fig. 7) and Mexico in South America (Nov. 13 to Dec. 2, 2020, Fig 8). The reason for $t_c = 12$ in Italy may be that Italy declared a national curfew and a state of emergency in the Lombardy region and three other regions on Nov. 5, 2020. Similarly, $t_c = 7$ in Mexico is due to that the centralized resettlement of victims caused by Hurricane Eta on Nov. 11, 2020 reduced people's social distance. Furthermore, for countries and regions with longer time span and more complex situations, the number of change points can be increased to make the simulation close to the actual situation. According to the previous studies, we believe that our Markov method can be generally applied to the development of COVID-19 in different countries or regions.

In what follows, we evaluate the control effect of non-drug intervention strategies. The non-drug intervention strategies can affect three parameters in the process of COVID-19 transmission: The initial number of active cases ($k$), the infection rate ($\lambda$), and the infection batch size ($d$). Note that the disappearing rate ($\mu$) does not change under the non-drug intervention strategies.

It is now generally accepted that there are four kinds of non-drug intervention strategies to limit the spread of COVID-19. The four strategies are as follows: Travel restrictions, early identification and isolation, increasing social distance, and exposure restrictions. In this study, we find the change point of key parameters within the observation period by means of real data. According to the comparison between the change point and the actual non-drug intervention strategies, we can see the control effect of intervention strategies. Note that the non-drug intervention strategies may affect the three parameters ($\lambda$, $d$, and $k$) at the same time, and the change of parameters has a certain delay period compared with the time of policy adoption.

Next, we analyze the effect of the non-drug intervention strategies on the control of COVID-19 transmission. We take a 20-day observation time scale and use the data of active cases within 20 days of COVID-19 transmission. Let $\theta$ and $\tilde{\theta}$ be two values of a key parameter in the transmission process of COVID-19. We denote by $E[N_\theta(t)]$ and $E[N_{\tilde{\theta}}(t)]$ the average numbers of COVID-19 patients corresponding to the two values $\theta$ and $\tilde{\theta}$, respectively. We define

$$\rho_{\theta,\tilde{\theta}} = \frac{E[N_\theta(t)]}{E[N_{\tilde{\theta}}(t)]},$$

which expresses the control effect of COVID-19 transmission when the key parameter of COVID-19 transmission is changed from $\theta$ to $\tilde{\theta}$.

In order to analyze the control effect, we choose South Korea as an example. The parameters have a change point, $t_c = 11$, before which the parameters are as follows: The average infection rate $\lambda = 0.062135947$, the average disappearing rate $\mu = 0.053096363$, the initial number of active cases $k = 1825$, and the weights ($r$) of the two groups $d_1^1 = 1$, $d_2^1 = 2$, equal 0.940 and 0.060, respectively. The parameters after $t_c$ are as follows: The average infection rate $\lambda = 0.09774732$, the average disappearing rate $\mu = 0.038994525$, and the weights ($r$) of the two groups $d_1^2 = 1$, $d_2^2 = 2$, equal 0.434 and 0.566, respectively. We used the above actual data of South Korea to analyze the control effect. To this end, we suppose some non-drug intervention strategies can change the parameters as follows: $\tilde{\lambda} = \frac{1}{2}\lambda$; $\tilde{k} = \left\lceil \frac{1}{2}k \right\rceil$, where $\lceil x \rceil$ denotes the minimum integer higher than $x$; $d_1^i = 1$ change to $\tilde{d}_1^i = 0$, $d_2^i = 2$ change to $\tilde{d}_2^i = 1$ for $i = 1, 2$; $\tilde{r}_1^1 = 0.470$, $\tilde{r}_2^1 = 0.530$, $\tilde{r}_1^2 = 0.217$, $\tilde{r}_2^2 = 0.783$ are changed from $r_1^1 = 0.940$, $r_2^1 = 0.060$, $r_1^2 = 0.434$, $r_2^2 = 0.566$, respectively. These parameters' decrease clearly reflects the effect of the non-drug intervention strategies on the control of COVID-19 transmission. From Fig. 9, it can be seen that the influence of the initial value of active cases, $k$, on the newly infected cases is smaller, compared to the explosive transmission process within the next periods; and the influences of the infection rate, $\lambda$, and the infection batch size, $d$, are significant. Therefore, it is better to reduce $\lambda$ and $d$, rather than $k$ in order to control COVID-19 transmission effectively.

Now, we take Egypt as another example to analyze such control effect. The parameters of transmission process in Egypt are as follows: $\lambda = 0.085434063$, $\mu = 0.045978143$, $k = 1903$, and the weights ($r$) of the two groups with $d_1^1=1$, $d_2^1=2$ equal 0.946 and 0.054, respectively. We also use some non-drug intervention strategies to change the parameters: $\tilde{\lambda} = \frac{1}{2}\lambda$; $\tilde{k} = \left\lceil \frac{1}{2}k \right\rceil$; $d_1 = 1$ changes to $\tilde{d}_1 = 0$, $d_2 = 1$ changes to $\tilde{d}_2 = 0$; $\tilde{r}_1 = 0.518$, $\tilde{r}_1 = 0.482$ are changed from $r_1 = 0.946$, $r_1 = 0.054$, respectively. From Fig. 10, it is found that the influence trend of the three parameters in Egypt on the transmission process is similar to that in South Korea.

To reduce the infection rate $\lambda$, the infection batch size $d$, and the initial value of active cases $k$, it is observed from Figures 9 and 10 that we need to further sufficiently adopt some non-drug intervention strategies, such as travel restrictions, community isolation, reducing parties, wearing masks, etc.

## Discussion

In this research, we develop a Markov system to model the COVID-19 transmission process. To the best of our knowledge, the transmission of COVID-19 has been mainly studied by SIR models and their extensions in the existing research. Our research represents the first time that the theory of Markov processes is used to study such a problem. That is, we propose a new direction in the study of COVID-19 transmission. We believe that the methodology and results given in this paper will open a new avenue to the study of COVID-19 transmission, and can motivate a series of promising future research on various new mutations of COVID-19.

One merit of our Markov process for the COVID-19 transmission is that it contains only four basic parameters: The infection rate $\lambda$, the disappearing rate $\mu$, the infection batch size $d$, and the number of initial active cases, $k$. On the one hand, by using these parameters we can describe the COVID-19 transmission process in details. On the other hand, these parameters can be used to evaluate the transmission process of COVID-19, and, therefore, they can be used to estimate the control effect of non-drug intervention strategies, and even the impact of vaccine application. Thus, these parameters act as a bridge between the intervention strategies and the COVID-19 transmission process. This connection can effectively provide guidance for decision making on using the non-drug intervention strategies.

The transmission of COVID-19 happens in a complex environment, where different patients may have different infection ability. It is difficult to use a single Markov process to describe such a problem. To better analyze this process, we divide the random system of COVID-19 into several groups and treat each group as an independent Markov process, as seen in Fig. 2. The transmission of COVID-19 within a country or a city can be described by means of combining multiple Markov processes of different groups. According to our empirical examples, we find that it is enough to use only two groups with different parameters to describe the real short-term change process of the countries and cities which have serious COVID-19 transmission. In addition, in the history of applications of Markov processes, our research is the first one to combine multiple Markov processes to describe and analyze a practical stochastic system.

In our study, we consider three values of the infection batch size, $d$: $d=0$,

$d=1$, and $d=2$, which represent three different levels of infection batch sizes. When there are two groups in a country or city, we have three pairs: $d_1=0, d_2=1$; $d_1=0, d_2=2$; and $d_1=1, d_2=2$. These three pairs represent low, medium and high levels of infection batch sizes in the country or city, respectively. In our simulation process, we find that the three infection batch sizes ($d=0$, $d=1$, and $d=2$) are sufficient for describing current COVID-19 transmission processes in the world. As the COVID-19 mutates fast, when necessary we can consider more values of $d$ (for example, $d=$ 0, 1, 2, 3, 4, 5) to express the more detailed levels of COVID-19 transmission.

The transmission intensity of COVID-19 in a country or city may change over time, because of the adoption of the non-drug intervention strategies. This can be seen from the change points of these parameters in the transmission process of COVID-19. These change points quantitatively evaluate the control effect of non-drug intervention strategies. The change points can be set in different groups, which enables our simulation to well reflect the short-term transmission processes of COVID-19 under non-drug intervention strategies.

Note that the simulation of the transmission process of COVID-19 by using our Markov system can fit the actual situation of COVID-19 very well, and our method can also be well applied to the prediction of transmission processes of COVID-19. Such a forecast can be developed in two types: (1) No intervention strategy. We focus on predicting the future transmission process using our simulation approach based on changing the value of the initial infection size $k$, while keeping the other three parameters ($\lambda$, $\mu$, and $d$) unchanged. (2) The non-drug intervention strategies are adopted. When there are non-drug intervention strategies are considered to implement, we can roughly forecast the trend of the resulted future transmission process. This trend can be used to help decision making on COVID-19 intervention strategies within a country/city. Our Markov system and associated simulation technique can be used to support controlling the transmission processes.

Currently, some countries have adopted the COVID-19 vaccines. Therefore, our future research efforts will be devoted to adapting the proposed approach to not only evaluate the effect of the vaccines on the prevention of COVID-19, but also predict the new transmission trend during the future process of COVID-19 mutation.

Table 1 Data on COVID-19 cases of New York State from Nov. 6 to 25, 2020

| Date | Confirmed cases | Daily confirmed cases | Disappearing cases | Daily disappearing cases | Active cases |
|---|---|---|---|---|---|
| Nov. 6 | 559161 | 3241 | 457569 | 721 | 101592 |
| Nov. 7 | 562577 | 3416 | 458342 | 773 | 104235 |
| Nov. 8 | 565929 | 3352 | 459035 | 693 | 106894 |
| Nov. 9 | 569508 | 3579 | 459873 | 838 | 109635 |
| Nov. 10 | 573465 | 3957 | 460601 | 728 | 112864 |
| Nov. 11 | 578076 | 4611 | 461267 | 666 | 116809 |
| Nov. 12 | 583133 | 5057 | 461817 | 550 | 121316 |
| Nov. 13 | 588605 | 5472 | 462469 | 652 | 126136 |
| Nov. 14 | 593767 | 5162 | 463161 | 692 | 130606 |
| Nov. 15 | 597394 | 3627 | 464257 | 1096 | 133137 |
| Nov. 16 | 601457 | 4063 | 465006 | 749 | 136451 |
| Nov. 17 | 606624 | 5167 | 465711 | 705 | 140913 |
| Nov. 18 | 611988 | 5364 | 466524 | 813 | 145464 |
| Nov. 19 | 617741 | 5753 | 467298 | 774 | 150443 |
| Nov. 20 | 623242 | 5501 | 467354 | 56 | 155888 |
| Nov. 21 | 628808 | 5566 | 469471 | 2117 | 159337 |
| Nov. 22 | 634035 | 5227 | 470455 | 984 | 163580 |
| Nov. 23 | 640356 | 6321 | 471737 | 1282 | 168619 |
| Nov. 24 | 645834 | 5478 | 472576 | 839 | 173258 |
| Nov. 25 | 651830 | 5996 | 473349 | 773 | 178481 |

Data source: https://github.com/CSSEGISandData/COVID-19.

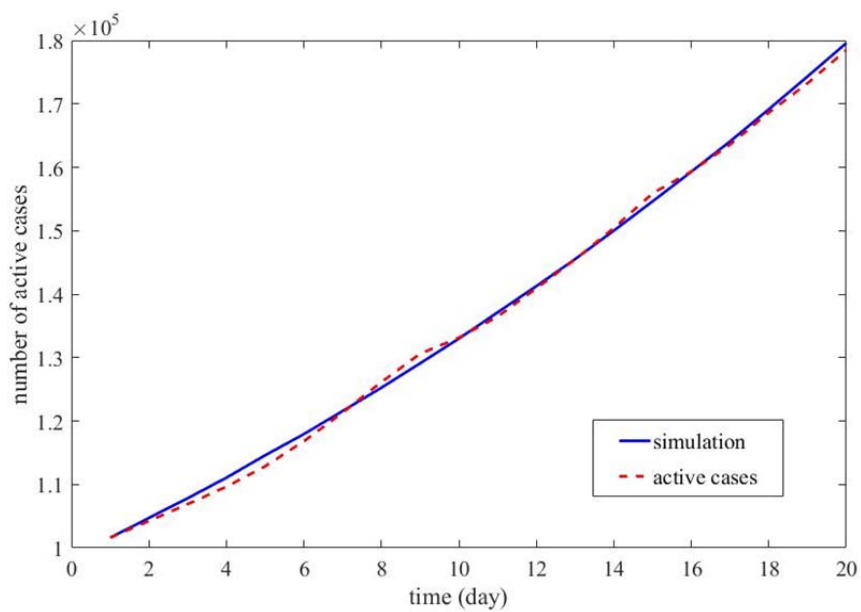

**Fig. 3 The growth process of active cases of New York State**

Table 2 Data on COVID-19 cases of India from Nov. 1 to 20, 2020

| Date | Confirmed cases | Daily confirmed cases | Disappearing cases | Daily disappearing cases | Active cases |
|---|---|---|---|---|---|
| Nov. 1 | 8229313 | 45231 | 7667405 | 53781 | 561908 |
| Nov. 2 | 8267623 | 38310 | 7726218 | 58813 | 541405 |
| Nov. 3 | 8313876 | 46253 | 7780089 | 53871 | 533787 |
| Nov. 4 | 8364086 | 50210 | 7836124 | 56035 | 527962 |
| Nov. 5 | 8411724 | 47638 | 7890951 | 54827 | 520773 |
| Nov. 6 | 8462080 | 50356 | 7945448 | 54497 | 516632 |
| Nov. 7 | 8507754 | 45674 | 7995089 | 49641 | 512665 |
| Nov. 8 | 8553657 | 45903 | 8043984 | 48895 | 509673 |
| Nov. 9 | 8591730 | 38073 | 8086465 | 42481 | 505265 |
| Nov. 10 | 8636011 | 44281 | 8141354 | 54889 | 494657 |
| Nov. 11 | 8683916 | 47905 | 8194622 | 53268 | 489294 |
| Nov. 12 | 8728795 | 44879 | 8244248 | 49626 | 484547 |
| Nov. 13 | 8773479 | 44684 | 8292760 | 48512 | 480719 |
| Nov. 14 | 8814579 | 41100 | 8335363 | 42603 | 479216 |
| Nov. 15 | 8845127 | 30548 | 8379649 | 44286 | 465478 |
| Nov. 16 | 8874290 | 29163 | 8420889 | 41240 | 453401 |
| Nov. 17 | 8912907 | 38617 | 8466102 | 45213 | 446805 |
| Nov. 18 | 8958483 | 45576 | 8515180 | 49078 | 443303 |
| Nov. 19 | 9004365 | 45882 | 8560571 | 45391 | 443794 |
| Nov. 20 | 9050597 | 46232 | 8610850 | 50279 | 439747 |

Data source: https://github.com/CSSEGISandData/COVID-19.

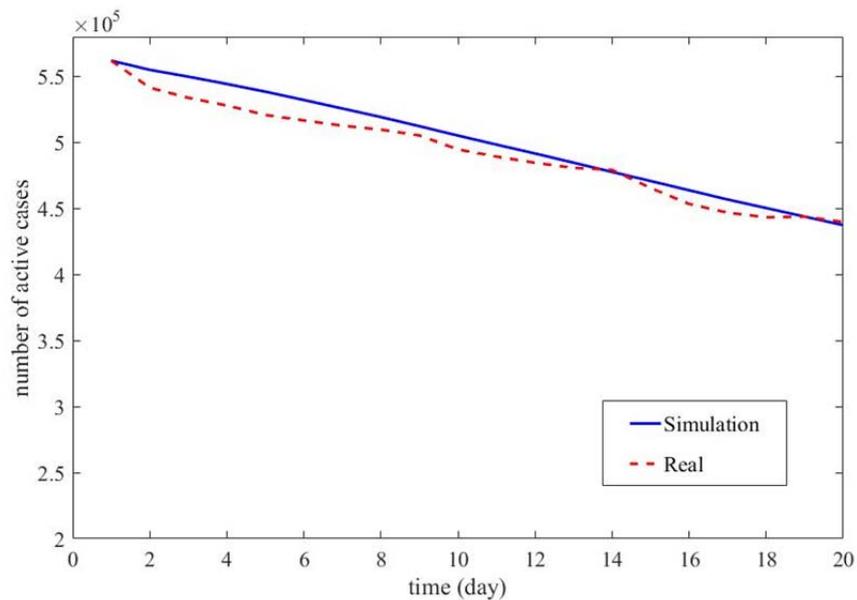

**Fig. 4 The growth process of confirmed cases of India.** According to the data of India from Nov.

1 to 20, 2020, the parameters of this period are obtained: The average infection rate ($\lambda$) is 0.088146484 and the average disappearing rate ($\mu$) is 0.101284201. $d_1 = 0$, $d_2 = 1$ and the weights ($r$) of $d_1$ and $d_2$ are 0.001 and 0.999, respectively. The initial number of active cases (k) at the beginning of our observation period (Nov. 1) is determined as 561908.

Table 3 Data on COVID-19 cases of Egypt from Nov. 1 to 20, 2020

| Date | Confirmed cases | Daily confirmed cases | Disappearing cases | Daily disappearing cases | Active cases |
|---|---|---|---|---|---|
| Nov. 1 | 107736 | 181 | 105833 | 115 | 1903 |
| Nov. 2 | 107925 | 189 | 105943 | 110 | 1982 |
| Nov. 3 | 108122 | 197 | 106070 | 127 | 2052 |
| Nov. 4 | 108329 | 207 | 106192 | 122 | 2137 |
| Nov. 5 | 108530 | 201 | 106335 | 143 | 2195 |
| Nov. 6 | 108754 | 224 | 106449 | 114 | 2305 |
| Nov. 7 | 108962 | 208 | 106594 | 145 | 2368 |
| Nov. 8 | 109201 | 239 | 106710 | 116 | 2491 |
| Nov. 9 | 109422 | 221 | 106819 | 109 | 2603 |
| Nov. 10 | 109654 | 232 | 106934 | 115 | 2720 |
| Nov. 11 | 109881 | 227 | 107067 | 133 | 2814 |
| Nov. 12 | 110095 | 214 | 107177 | 110 | 2918 |
| Nov. 13 | 110319 | 224 | 107276 | 99 | 3043 |
| Nov. 14 | 110547 | 228 | 107388 | 112 | 3159 |
| Nov. 15 | 110767 | 220 | 107499 | 111 | 3268 |
| Nov. 16 | 111009 | 242 | 107644 | 145 | 3365 |
| Nov. 17 | 111284 | 275 | 107769 | 125 | 3515 |
| Nov. 18 | 111613 | 329 | 107916 | 147 | 3697 |
| Nov. 19 | 111955 | 342 | 108072 | 156 | 3883 |
| Nov. 20 | 112318 | 363 | 108206 | 134 | 4112 |

Data source: https://github.com/CSSEGISandData/COVID-19.

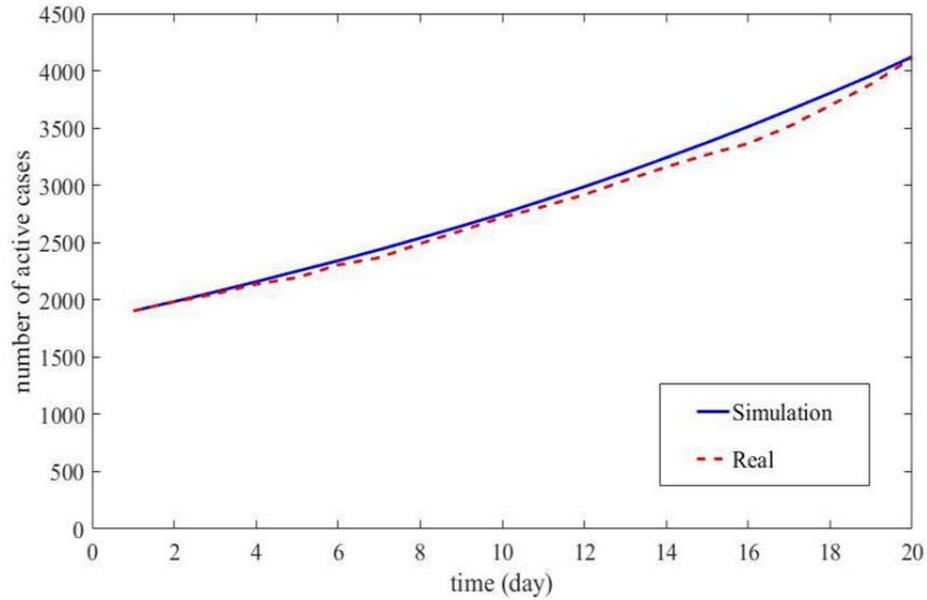

**Fig. 5 The growth process of confirmed cases of Egypt.** According to the data of Egypt from Nov. 1 to 20, 2020, the four parameters within this period are obtained: The average infection rate ($\lambda$) is 0.085434063 and the average disappearing rate ($\mu$) is 0.045978143. $d_1$=1, $d_2$=2, and the weights ($r$) of $d_1$ and $d_2$ are 0.946 and 0.054, respectively. The initial number of active cases ($k$) at the beginning of our observation period (Nov. 1) is determined as 1903.

Table 4 Data on COVID-19 cases of South Korea from Nov. 2 to 21, 2020

| Date | Confirmed cases | Daily confirmed cases | Disappearing cases | Daily disappearing cases | Active cases |
|---|---|---|---|---|---|
| Nov. 2 | 26807 | 75 | 24982 | 119 | 1825 |
| Nov. 3 | 26925 | 118 | 25090 | 108 | 1835 |
| Nov. 4 | 27050 | 125 | 25210 | 120 | 1840 |
| Nov. 5 | 27195 | 145 | 25297 | 87 | 1898 |
| Nov. 6 | 27284 | 89 | 25387 | 90 | 1897 |
| Nov. 7 | 27427 | 143 | 25446 | 59 | 1981 |
| Nov. 8 | 27553 | 126 | 25509 | 63 | 2044 |
| Nov. 9 | 27653 | 100 | 25645 | 136 | 2008 |
| Nov. 10 | 27799 | 146 | 25753 | 108 | 2046 |
| Nov. 11 | 27942 | 143 | 25891 | 138 | 2051 |
| Nov. 12 | 28133 | 191 | 26025 | 134 | 2108 |
| Nov. 13 | 28338 | 205 | 26128 | 103 | 2210 |
| Nov. 14 | 28546 | 208 | 26184 | 56 | 2362 |

| Nov. 15 | 28769 | 223 | 26253 | 69 | 2516 |
| Nov. 16 | 28998 | 229 | 26354 | 101 | 2644 |
| Nov. 17 | 29311 | 313 | 26469 | 115 | 2842 |
| Nov. 18 | 29654 | 343 | 26596 | 127 | 3058 |
| Nov. 19 | 30017 | 363 | 26764 | 168 | 3253 |
| Nov. 20 | 30403 | 386 | 26868 | 104 | 3535 |
| Nov. 21 | 30733 | 330 | 26971 | 103 | 3762 |

Data source: https://github.com/CSSEGISandData/COVID-19.

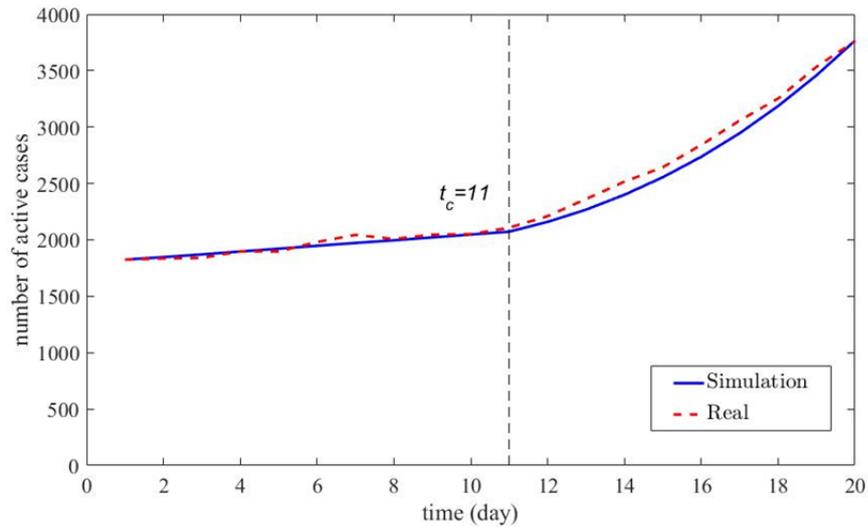

**Fig. 6 The growth process of confirmed cases of South Korea**

Table 5 Data on COVID-19 cases of Italy from Nov. 11 to 30, 2020

| Date | Confirmed cases | Daily confirmed cases | Disappearing cases | Daily disappearing cases | Active cases |
| --- | --- | --- | --- | --- | --- |
| Nov. 11 | 1028424 | 32961 | 415066 | 9713 | 613358 |
| Nov. 12 | 1066401 | 37977 | 431347 | 16281 | 635054 |
| Nov. 13 | 1107303 | 40902 | 443377 | 12030 | 663926 |
| Nov. 14 | 1144552 | 37249 | 456117 | 12740 | 688435 |
| Nov. 15 | 1178529 | 33977 | 466039 | 9922 | 712490 |
| Nov. 16 | 1205881 | 27352 | 488097 | 22058 | 717784 |
| Nov. 17 | 1238072 | 32191 | 504262 | 16165 | 733810 |
| Nov. 18 | 1272352 | 34280 | 529184 | 24922 | 743168 |
| Nov. 19 | 1308528 | 36176 | 546857 | 17673 | 761671 |
| Nov. 20 | 1345767 | 37239 | 568591 | 21734 | 777176 |
| Nov. 21 | 1380531 | 34764 | 588785 | 20194 | 791746 |
| Nov. 22 | 1408868 | 28337 | 602921 | 14136 | 805947 |
| Nov. 23 | 1431795 | 22927 | 634946 | 32025 | 796849 |

| Nov. 24 | 1455022 | 23227 | 656636 | 21690 | 798386 |
| Nov. 25 | 1480874 | 25852 | 689177 | 32541 | 791697 |
| Nov. 26 | 1509875 | 29001 | 714030 | 24853 | 795845 |
| Nov. 27 | 1538217 | 28342 | 750324 | 36294 | 787893 |
| Nov. 28 | 1564532 | 26315 | 775224 | 24900 | 789308 |
| Nov. 29 | 1585178 | 20646 | 789407 | 14183 | 795771 |
| Nov. 30 | 1601554 | 16376 | 813083 | 23676 | 788471 |

Data source: https://github.com/CSSEGISandData/COVID-19.

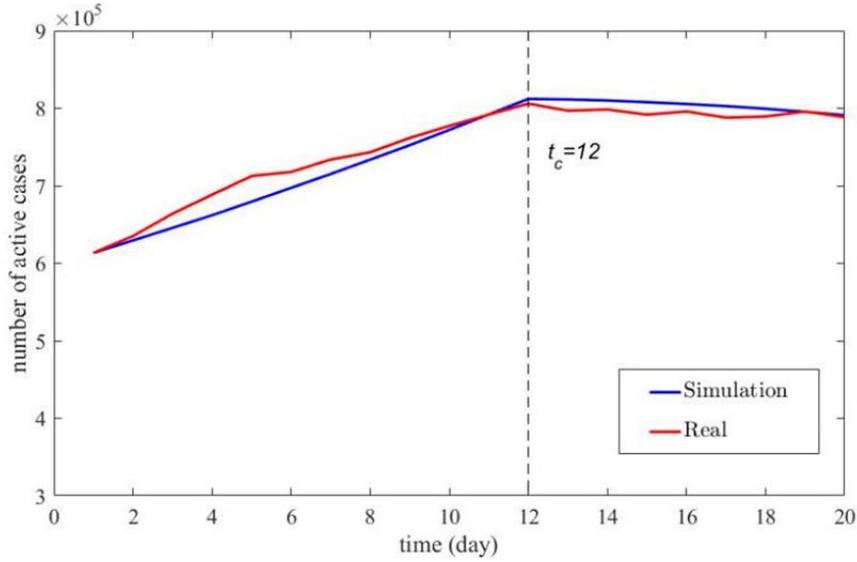

**Fig. 7 The growth process of confirmed cases of Italy.** The parameters of the observation period in Italy were as follows: $t_c = 12$. The parameters before $t_c$ are as follows: The average infection rate is 0.049487386 ($\lambda$) and the average disappearing rate ($\mu$) is 0.023181119. The initial number of active cases ($k$) at the beginning of our observation period (Nov. 11) is determined as 613358. We then simulated the development of active cases in Italy during this period. In order to obtain simulated results close to the real data, we infer that the weights ($r$) of the two groups $d_1^1 = 1$, $d_2^1 = 2$ equal 0.999 and 0.001, respectively. The parameters after $t_c$ are as follows: The average infection rate ($\lambda$) is 0.030904889 and the average disappearing rate ($\mu$) is 0.031409968. the weights ($r$) of the two groups $d_1^2 = 0$, $d_2^2 = 1$ equal 0.550 and 0.450, respectively.

Table 6 Data on COVID-19 cases of Mexico from Nov. 11 to 30, 2020

| Date | Confirmed cases | Daily confirmed cases | Disappearing cases | Daily disappearing cases | Active cases |
|---|---|---|---|---|---|
| Nov. 13 | 997393 | 5558 | 838964 | 5711 | 158429 |
| Nov. 14 | 1003253 | 5860 | 843620 | 4656 | 159633 |
| Nov. 15 | 1006522 | 3269 | 848732 | 5112 | 157790 |
| Nov. 16 | 1009396 | 2874 | 853061 | 4329 | 156335 |
| Nov. 17 | 1011153 | 1757 | 856977 | 3916 | 154176 |
| Nov. 18 | 1015071 | 3918 | 861553 | 4576 | 153518 |
| Nov. 19 | 1019543 | 4472 | 866465 | 4912 | 153078 |
| Nov. 20 | 1025969 | 6426 | 871551 | 5086 | 154418 |
| Nov. 21 | 1032688 | 6719 | 872101 | 550 | 160587 |
| Nov. 22 | 1041875 | 9187 | 880780 | 8679 | 161095 |
| Nov. 23 | 1049358 | 7483 | 886619 | 5839 | 162739 |
| Nov. 24 | 1060152 | 10794 | 894255 | 7636 | 165897 |
| Nov. 25 | 1070487 | 10335 | 901634 | 7379 | 168853 |
| Nov. 26 | 1078594 | 8107 | 907823 | 6189 | 170771 |
| Nov. 27 | 1089998 | 11404 | 908431 | 608 | 181567 |
| Nov. 28 | 1101403 | 11405 | 909040 | 609 | 192363 |
| Nov. 29 | 1107071 | 5668 | 924052 | 15012 | 183019 |
| Nov. 30 | 1113543 | 6472 | 929526 | 5474 | 184017 |
| Dec. 1 | 1122362 | 8819 | 936582 | 7056 | 185780 |
| Dec. 2 | 1133613 | 11251 | 944132 | 7550 | 189481 |

Data source: https://github.com/CSSEGISandData/COVID-19.

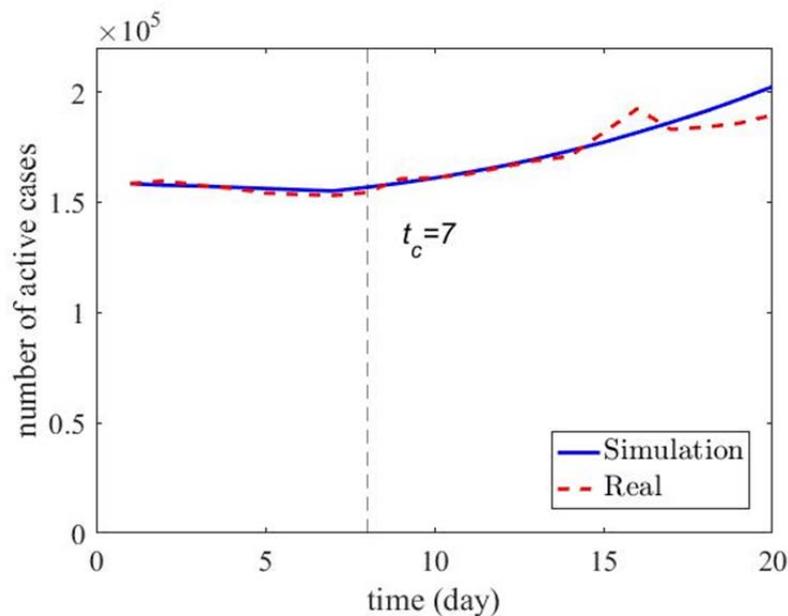

**Fig. 8 The growth process of confirmed cases of Mexico.** The parameters of the observation period in Mexico were as follows: $t_c$=7. The parameters before $t_c$ are as follows: The average

infection rate ($\lambda$) is 0.027329732 and the average disappearing rate ($\mu$) is 0.030691817. The initial number of active cases ($k$) at the beginning of our observation period (Nov. 13) is determined as 158429. We then simulated the development of active cases in Mexico during this period. In order to obtain simulated results close to the real data, we infer that the weights ($r$) of the two groups $d_1^1=0$, $d_2^1=1$ equal 0.020 and 0.980, respectively. The parameters after $t_c$ are as follows: The average infection rate ($\lambda$) is 0.051140063 and the average disappearing rate ($\mu$) is 0.034605204. the weights ($r$) of the two groups $d_1^2=1$, $d_2^2=2$ equal 0.500 and 0.500, respectively.

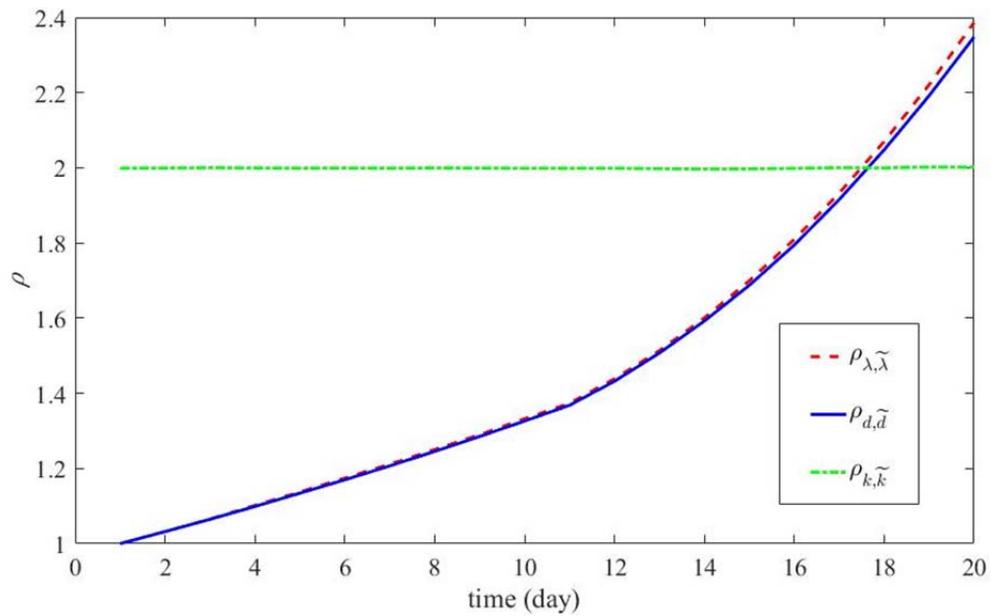

**Fig 9 The changing curve of $\rho$ in Korea when $\lambda$, $d$ and $k$ are different.**

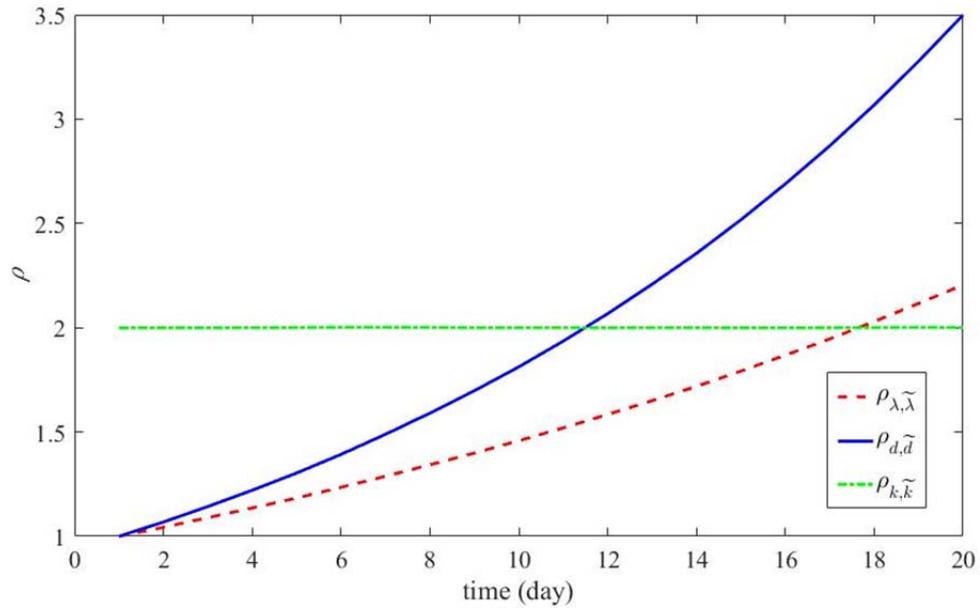

**Fig 10 The changing curve of** $\rho$ **in Egypt when** $\lambda$, $d$ **and** $k$ **are different.**